\documentclass[prd,twocolumn,preprintnumbers]{revtex4}

\usepackage{amsmath}
\usepackage{epsfig}
\usepackage{graphicx}
\usepackage{color}
\usepackage[normalem]{ulem}
 \usepackage{url}
 \usepackage{float}
\usepackage[breaklinks, plainpages=false, colorlinks=true, anchorcolor=cyan, linkcolor=red, citecolor=cyan, urlcolor=magenta, bookmarks=false]{hyperref}

\usepackage[caption=false]{subfig}

\begin{document}
\renewcommand{\thefigure}{\arabic{figure}}
\setcounter{figure}{0}

 \def\I{{\rm i}}
 \def\E{{\rm e}}
 \def\D{{\rm d}}

\bibliographystyle{apsrev}

\title{Low Latency Detection of Massive Black Hole Binaries}

\author{Neil J. Cornish}
\affiliation{eXtreme Gravity Institute, Department of Physics, Montana State University, Bozeman, Montana 59717, USA}

\begin{abstract} 
The next decade is expected to see the launch of one or more space based gravitational wave detectors: the European lead Laser Interferometer Space Antenna (LISA); and one or more Chinese mission concepts, Taiji and TianQin. One of the primary scientific targets for these missions are the mergers of black holes with masses between $10^3 M_\odot$ and $10^8 M_\odot$. These systems may produce detectable electromagnetic signatures in additional to gravitational waves due to the presence of gas in mini-disks around each black hole, and a circumbinary disk surrounding the system. The electromagnetic emission may occur before, during and after the merger. In order to have the best chance of capturing all phases of the emission it is imperative that the gravitational wave signals can be detected in low latency, and used to produce reliable estimates for the sky location and distance to help guide the search for counterparts. The low latency detection also provides a starting point for the ``global fit'' of the myriad signals that are simultaneously present in the data. Here a low latency analysis pipeline is presented that is capable of analyzing months of data in just a few hours using a laptop from the last decade. The problem of performing a global fit is avoided by whitening out the bright foreground produced by nearby galactic binaries. The performance of the pipeline is illustrated using simulated data from the LISA Data Challenge.
\end{abstract}

\maketitle

\section{Introduction}

One of the primary science goals for a future space based gravitational wave interferometer\, such as LISA~\cite{Baker:2019nia} Taiji~\cite{taiji}, and TianQin~\cite{TianQin}, is detecting the signals from massive black hole mergers with masses in the range $10^3 M_\odot -10^8 M_\odot$. In contrast with the stellar origin black holes detected by ground based interferometers, where the mergers are thought to occur in a matter-free environments (though see Ref.~\cite{McKernan_2019} for a possible counter example), massive black hole mergers are expected to occur in gas rich environments that can result in the production of electromagnetic counterpart signals: for a recent review see Ref.~\cite{Bogdanovic:2021aav}. Low latency detection of these signals is crucial for finding electromagnetic counterparts~\cite{DalCanton:2019wsr}. Low latency detection will also form the first stage of the {\em Global Fit}~\cite{Cornish:2005qw,Littenberg:2020bxy} of the many thousands of overlapping signals expected to be found in the low frequency band probed by space based interferometers. For concreteness, the discussion will focus on the LISA detector, and use simulated data from the LISA Data Challenge ({\tt https://lisa-ldc.lal.in2p3.fr}), but the approach would work equally well for the Taiji detector. Some modifications would need to be made for the search to be used for the TianQin detector due to its faster orbital motion.

Massive black hole binaries are relatively easy to detect due to their often large signal-to-noise ratios (SNRs) and short duration within the sensitive frequency band of a LISA-like detector. Binary merger signals evolve through a long inspiral phase followed by a relatively brief merger and ringdown phase. The time spent in-band depends on the mass of the system, in particular on the combination of component masses $m_1,m_2$ that is known as the chirp mass, ${\cal M} = (m_1 m_2)^{3/5}/(m_1+m_2)^{1/3}$. Low chirp mass systems, such as stellar origin black hole binaries with component masses in the tens of solar masses, and extreme mass ratio inspirals (EMRIs), made up of a stellar remnant and a massive black hole, stay in-band for years or decades, slowly accumulating signal-to-noise. In contrast, massive black hole binaries have large chirp masses, and accumulate most of their signal-to-noise in a matter of weeks or days. Consequently, the majority of massive black hole binaries can be detecting using short data segments. The low latency search described here takes advantage of the short duration of the massive black hole binary signals by working with short (roughly month long) segments of data. 

In contrast to the grid-based searches used in LIGO-Virgo analyses~\cite{LIGOScientific:2019hgc}, the low latency search described here uses a multi-stage stochastic search based on a population Markov Chain Monte Carlo algorithm. The first step is to produce an estimate for the power spectral density of the noise. This is done using a variant of the wavelet de-noising and spline-line model that has been developed for low latency analyses of LIGO-Virgo data~\cite{Cornish:2021wxy}. The search then proceeds in three stages, using a sequence of likelihood functions. In the first two stages some of the system parameters are maximized over, rather than marginalized over. Because the LISA detector is effectively stationary during the short data segments being analyzed, the first stage of the search uses the familiar time, phase and amplitude maximized likelihood function from the LIGO-Virgo grid searches (see section 8.2 of Ref.~\cite{LIGOScientific:2019hgc}). The second stage of the analysis incorporates the detector motion and searches over sky location using a generalized F-statistic likelihood function~\cite{Jaranowski:1998qm,Cornish:2006ms}. The final stage of the analysis is fully Bayesian, and uses the heterodyning approach~\cite{Cornish:2010kf,Cornish:2021lje} to speed-up the calculation of the full likelihood by a factor of ten thousand. 

The paper begins with a description of the simulated data used in the analysis. This is followed by a step-by-step description of the low latency analysis pipeline, illustrated with examples from analyzing the simulated data. The paper concludes with a discussion of the steps needed to handle more realistic data sets, and eventually, the real data.

\section{Overview of the simulated data}

The milli-Hertz gravitational wave sky is expected to be populated by myriad sources, including millions of compact galactic binaries (predominantly detached white dwarf binaries); hundreds of massive black hole binaries with component masses in the range $10^3 M_\odot \rightarrow 10^8 M_\odot$; thousands of extreme mass ratio binaries made up of stellar remnants and massive black holes; and thousands of stellar mass black hole binaries that will eventually merge in the frequency band covered by ground based interferometers. More exotic signals, such as burst from hitherto unknown astrophysical systems, and stochastic signals from the early Universe may also be present in the data. The signals from these systems will be long lived, resulting in millions of signals being simultaneously being present in the data collected by space based interferometers. To properly account for the overlap between the signals a {\em Global Fit}~\cite{Cornish:2005qw,Littenberg:2020bxy} is required that simultaneously models all the resolvable signals, in addition to modeling the instrument noise and accounting for gaps and disturbances in the data. Because of the large dimensionality of the models used in the {\em Global Fit}, the full analysis will be computationally intensive.

\begin{figure}[htp]
\includegraphics[width=0.48\textwidth]{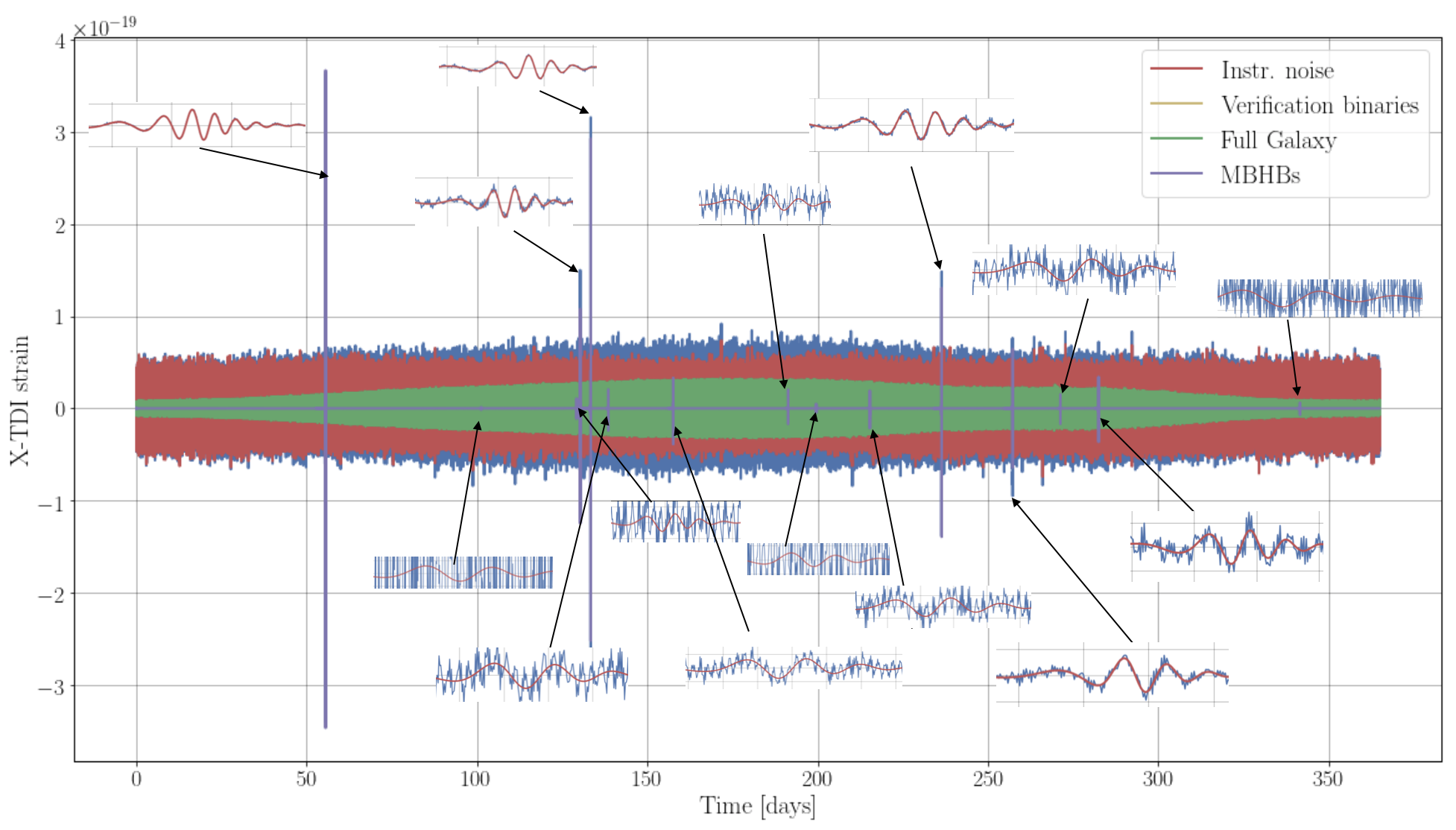} 
\caption{\label{fig:sangriatdi} Simulated X-Channel TDI data produced for the {\em Sangria} round of the LISA Data Challenge. The simulated data covers a little over one year, and includes colored instrument noise, signals from millions of galactic binaries, and fifteen massive black hole binary mergers. Figure courtesy of Stanislav Babak.}
\end{figure}

To help prepare for the rich data sets expected from the LISA mission, a series of data challenges have been carried out, starting with the Mock LISA Data Challenges~\cite{Arnaud:2006gm,Babak:2008aa,Babak:2009cj} in the 2000's, and resumed more recently as the LISA Data Challenge. The plan is to start with relatively simple challenges and to eventually build up to more realistic challenges that cover a full range of sources and instrumental complications. The challenges are rather whimsically named after adult beverages, with the perceived difficulty of the challenge indicated by the proof level of the libation. In this study the {\em Sangria} data set is used to demonstrate the performance of a prototype low latency search and characterization pipeline for massive black hole binaries. The {\em Sangria} data set includes millions of signals from a population synthesis model for galactic binaries, in addition to signals from fifteen massive black hole binaries drawn from an astrophysical population model. The data covers a roughly one year span, and includes simulated stationary and Gaussian noise. The {\em Sangria} data is free of gaps or other disturbances (the {\em Spritz} data set is the first to include gaps and glitches, but includes far fewer signals).

Figure~\ref{fig:sangriatdi} shows the  X-Channel Time Delay Interferometry data broken out into its various contributions. The galactic binary signals are modulated by the motion of the detector during the year, and from the perspective of a low latency search for massive black hole binaries, the galactic binaries can be thought of as a source of non-stationary and non-Gaussian noise. The massive black hole binary signals span a range of amplitudes, including some very loud signals that are clearly visible in the raw time domain data, and others that are buried in the instrumental and galactic foreground noise. The black hole binaries were simulated using the IMRPhenomD~\cite{Khan:2015jqa} waveform model, which covers just the dominant harmonic of spin-aligned (non-precessing) systems. Future challenges will employ more sophisticated waveform models that allow for spin-precession and multiple harmonics.

\section{Low Latency Search Pipeline}

Low latency means something different in the milli-Hertz frequency range covered by the LISA detector, as opposed to the kilo-Hertz range covered by the LIGO, Virgo, GEO and Kagra detectors. The timescales involved scale inversely with the frequencies, so while low-latency for ground based detectors is measured in seconds~\cite{canton2020realtime,Adams_2016}, low latency for space based detectors is measured in days. Indeed, the expectation is that the LISA data will only be transmitted to Earth every few days, with the possibility of more frequent transmissions if an interesting merger is imminent. The latter scenario depends on the system having been picked up days or weeks prior to merger.

Here we will take low-latency to imply the ability to analyze several months of data in a few hours with modest computing resources. The algorithm described here handily beats that goal. The search combines an improved version of the algorithm developed for the earlier {\em Radler} data challenge~\cite{Cornish:2020vtw}, with elements of the {\tt QuickCBC} algorithm~\cite{Cornish:2021wxy} that was developed to perform rapid parameter inference ground based interferometers.

\begin{figure}[htp]
\includegraphics[width=0.48\textwidth]{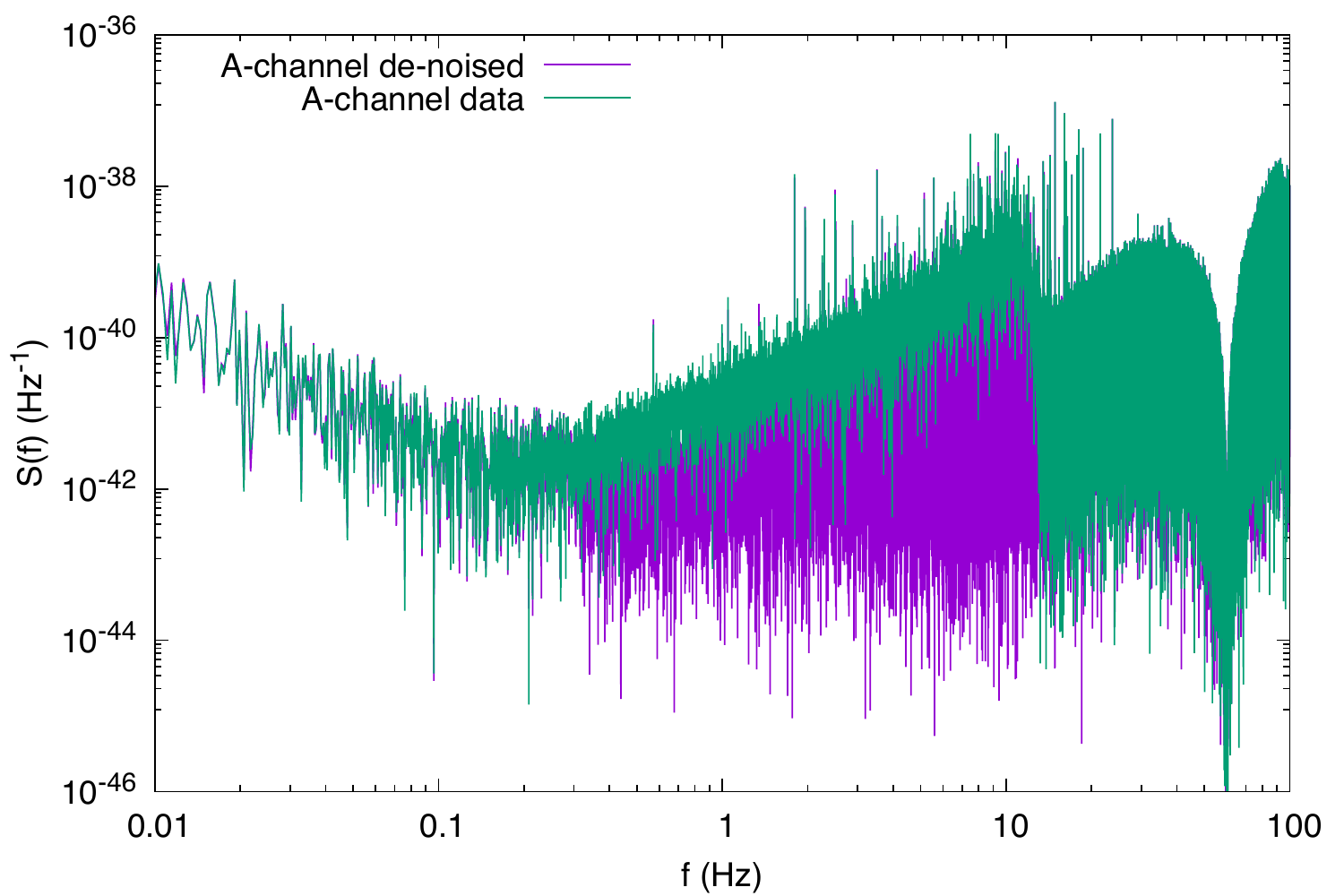} 
\caption{\label{fig:psd} Periodograms}
\end{figure}

The analysis starts with short segments of data, typically of order one month. The segments are overlapped by half their duration to provide robustness against edge effects that can occur due to window functions that are applied before Fourier transforming the data. Power spectra for each segment are estimated using the wavelet based algorithm from {\tt QuickCBC}~\cite{Cornish:2021wxy}. A key element of this approach is wavelet de-noising, which removes loud non-Gaussian and non-stationary features from the data that would otherwise distort the spectral estimate. In the case of LIGO/Virgo, the loud transients are mostly instrumental in origin, but with {\em Sangria} data set, it is the massive black holes. Figure~\ref{fig:psd} shows periodograms of the A-channel TDI data for month 2 of the {\em Sangria} training data, both before and after the wavelet de-noising. The presence of a very loud binary black hole merger during that month significantly distorts the raw periodogram.

\begin{figure}[htp]
\includegraphics[width=0.48\textwidth]{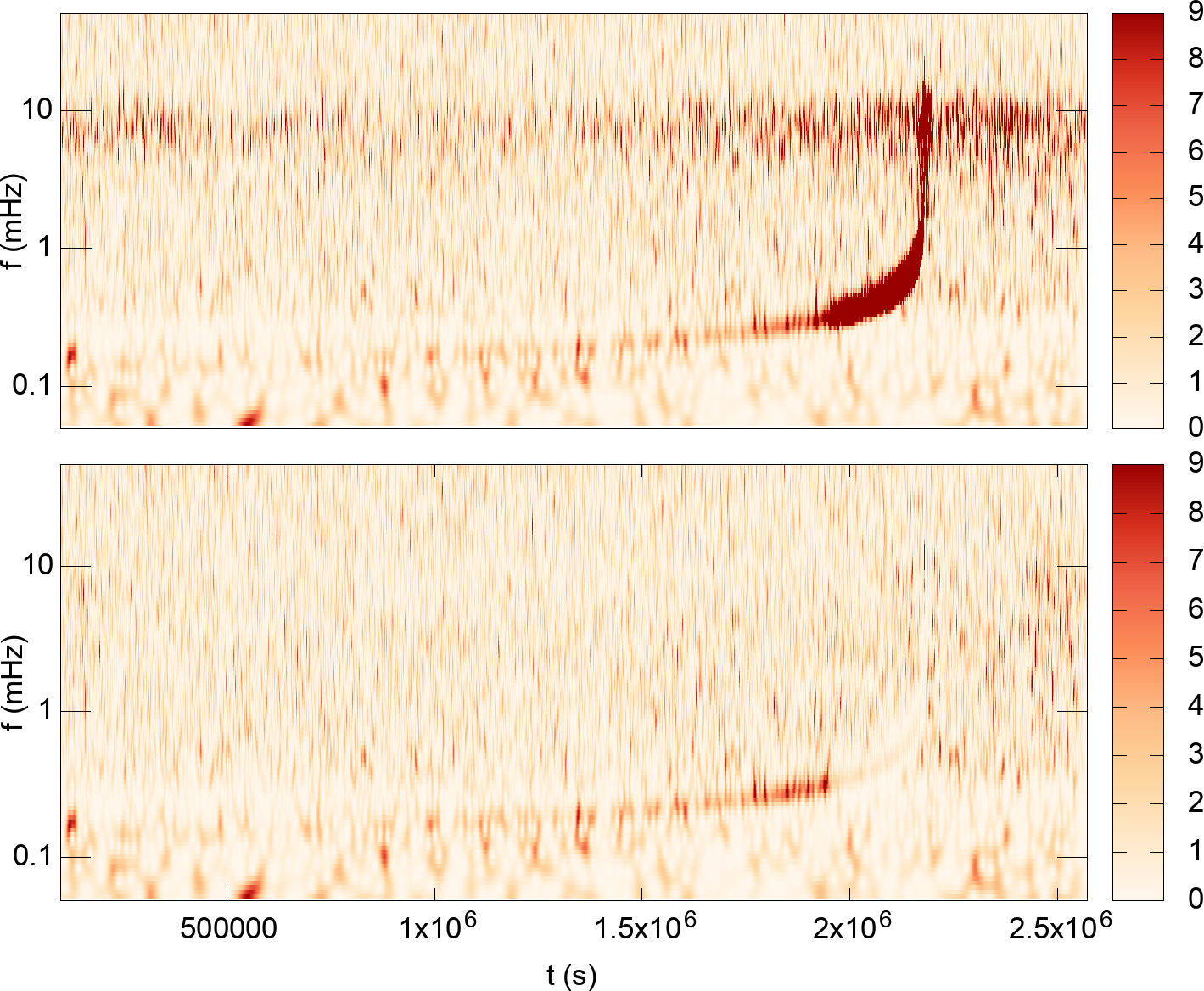} 
\caption{\label{fig:denoise} Wavelet de-noising applied to the A-channel TDI data for month 2 of the {\em Sangria} training data.}
\end{figure}

Wavelet de-nosing for the month 2 data segment is illustrated in Figure~\ref{fig:denoise}. Here the data has been whitened by the final spectral estimate and shown as a time-frequency map. While the wavelet de-noising does not remove all of the black hole signal, it removes enough to yield a reliable spectral estimate. Note that the color scale is clipped at maximum of nine, corresponding to a three-sigma excess in amplitude. The majority of the pixels that are removed by the wavelet denoising are orders of magnitude louder than that.

\begin{figure}[htp]
\includegraphics[width=0.48\textwidth]{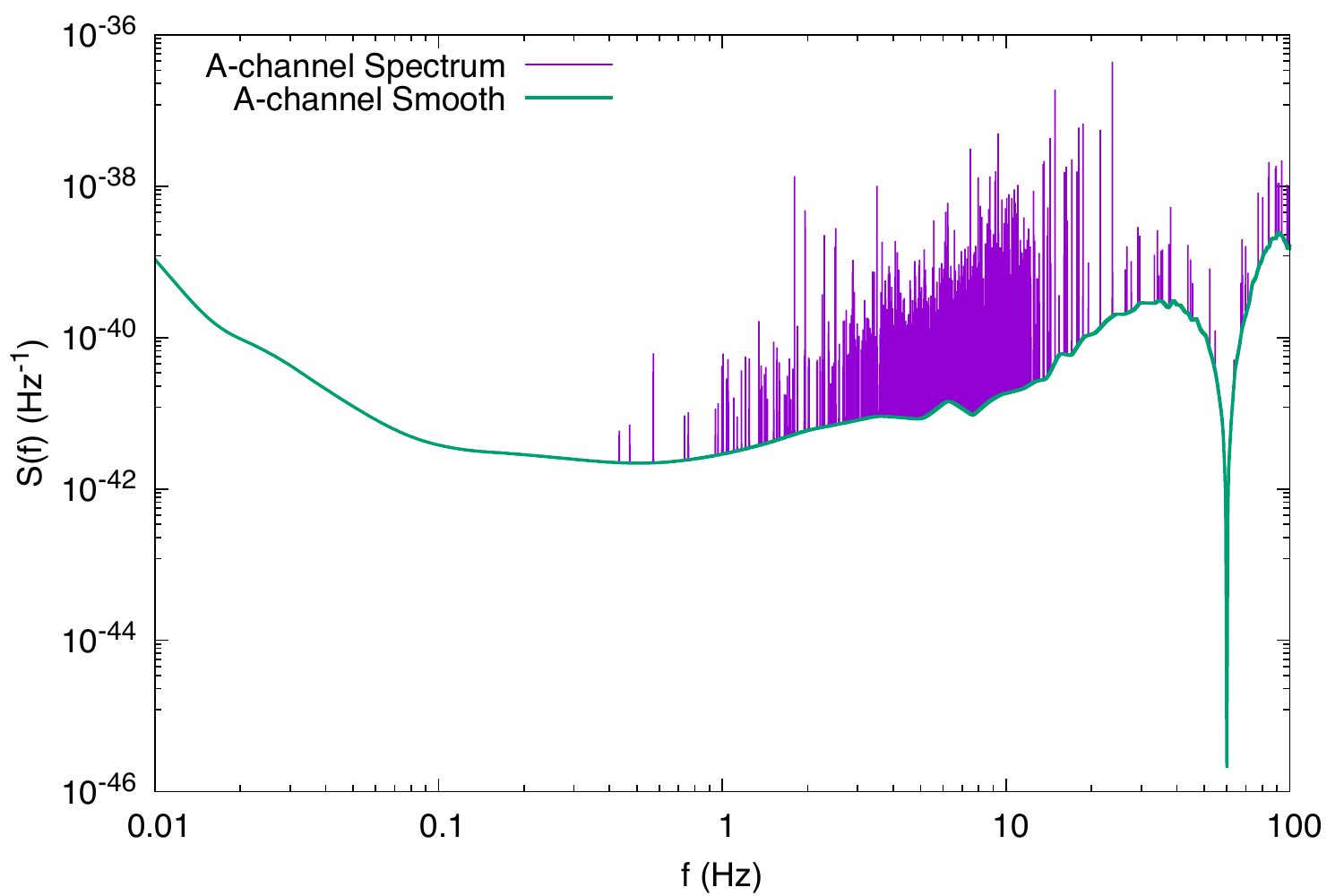} 
\caption{\label{fig:psdmodel} PSD model}
\end{figure}

The power spectrum is modeled using the fixed dimension version of the {\tt BayesLine} model~\cite{PhysRevD.91.084034} employed in the {\tt QuickCBC} pipeline. A cubic spline is used to model the smooth part of the spectrum, and Lorentzians are uses to model sharp spectral features. In contrast to the LIGO/Virgo application, where the lines are due to instrumental effects such as the suspension system, the lines seen in the spectral model shown in Figure~\ref{fig:psdmodel} are caused by bright white dwarf binaries. In other words, rather than solving for and subtracting the bright galactic binaries, as is done in the {\em Global Fit}, here the foreground is simply whitened away. Quieter galactic binaries, which will form the galactic confusion noise, are taken care of by the smooth spline component.

With a model for the power spectrum in place, the next step is to perform a Parallel Tempered Markov Chain Monte Carlo (PTMCMC) based search of the data. As described in Ref.~\cite{Cornish:2020vtw} the search works through a sequence of steps, each using a different likelihood function. The initial stage of the search uses a likelihood function that maximizes over merger time, and the overall amplitude and phase in each data channel. This likelihood ignores the sky location, distance and orientation of the source (inclination, polarization). It is very similar to the likelihoods used in grid-based searches of LIGO-Virgo data.  The justification for being able to use such a simple likelihood is as follows: The LISA instrument response can be written as~
\cite{Cornish:2020vtw}
\begin{equation}
h(f) = \left(F_+ A_+ + i F_\times A_\times \right) {\cal A}(f) e^{i \Phi(f)} \, ,
\end{equation}
where ${\cal A}(f)$ and $\Phi(f)$ are the intrinsic amplitude and phase of the signal (given here by the IMRPhenomD~\cite{Khan:2015jqa} waveform model), and $F_+(\theta,\phi,\psi,t(f),f)$, $F_\times(\theta,\phi,\psi,t(f),f)$ are the complex antenna patterns for a given channel. In the low frequency or long wavelength limit, where the wavelength of the gravitational wave is larger than the detector arms, the antenna patterns are real, and assume the familiar form used in LIGO-Virgo analyses~\cite{LIGOScientific:2019hgc}. The quantities $A_+ = (1+\cos^2\iota)/2$, $A_\times = \cos\iota$ depend on the inclination of the binary orbit with respect to the line of sight. For a precessing binary, both the polarization angle $\psi$, and inclination angle $\iota$, will vary with time (and hence frequency via the time-frequency mapping $t(f))$. We can define the polarization amplitude and phase as~\cite{Cutler:1997ta}
\begin{eqnarray}
A_p &=& \vert F_+ A_+ + i F_\times A_\times  \vert \nonumber \\
\phi_p &=& {\rm arg} \left(F_+ A_+ + i F_\times A_\times \right) \, ,
\end{eqnarray}
and write
\begin{equation}
h(f) = A_p(f) {\cal A}(f) e^{i (\Phi(f) + \phi_p(f))} \, .
\end{equation}
So far no approximations have been made. The approximation comes in treating these phases as constant over the segment. The maximized likelihood used in the first stage of the analysis returns a best fit value for $A_p$ and $\phi_p$ is each channel.

\begin{figure}[htp]
\includegraphics[width=0.48\textwidth]{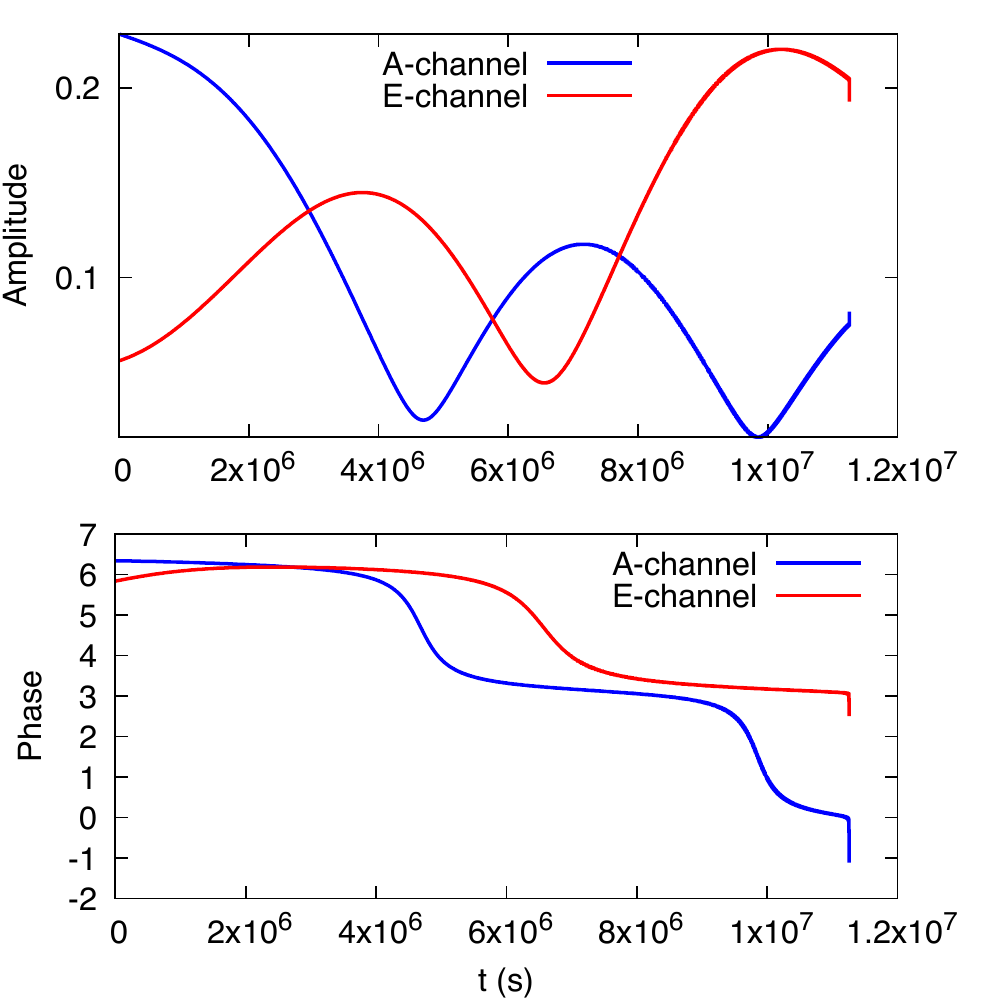} 
\caption{\label{fig:pol} Evolution of the amplitude and polarization phase for Source 4 (Detector frame masses $m_1 =9.16 \times 10^5$,  $m_2 = 7.02\times 10^5$, merger time $t_c= 1.125857 \times 10^7$).}
\end{figure}

Figure~\ref{fig:pol} shows the time evolution of the polarization amplitude $A_p(t)$ and phase $\phi_p(t)$ for source 4 from the {\em Sangria} data set (the source numbering follows the order in which the systems merge). In most one month segments, both the polarization phase and amplitude are slowly varying, and approximating them as constant in the first stage of the search does not result in a large loss of signal-to-noise. Note that the rapid evolution seen right around merger is not due to the motion of the detector, but rather, is due to the frequency dependence of the antenna response functions as the signal moves rapidly from the low frequency regime to the high frequency regime where finite armlength effects become important.

The first stage of the search quickly locks in on the signal from any massive black hole binary that might be present in the data segment. If there are multiple detectable signals, the search typically locks onto the loudest signal first. Following the first stage of the search the detector frame masses and merger time are well determined and the search moves to the second stage, which uses a F-statistic~\cite{Jaranowski:1998qm,Cornish:2006ms} likelihood to further refine the solution and to find the sky location and distance to the source. The F-statistic likelihood uses the full instrument response, and divides the waveform template into four non-orthogonal filters that are used to simultaneously maximize over distance $D_L$, inclination $\iota$, polarization $\psi$ and the merger phase $\phi_c$. The F-statistic likelihood references the merger time to the time at the guiding center of the LISA constellation, which also corresponds to the time returned by the maximization routine in the first stage of the search. The detector frame time is then mapped to the Barycenter merger time, $t_c$, using the sky location of the source and the position of the detector at merger.

\begin{figure}[htp]
\includegraphics[width=0.48\textwidth]{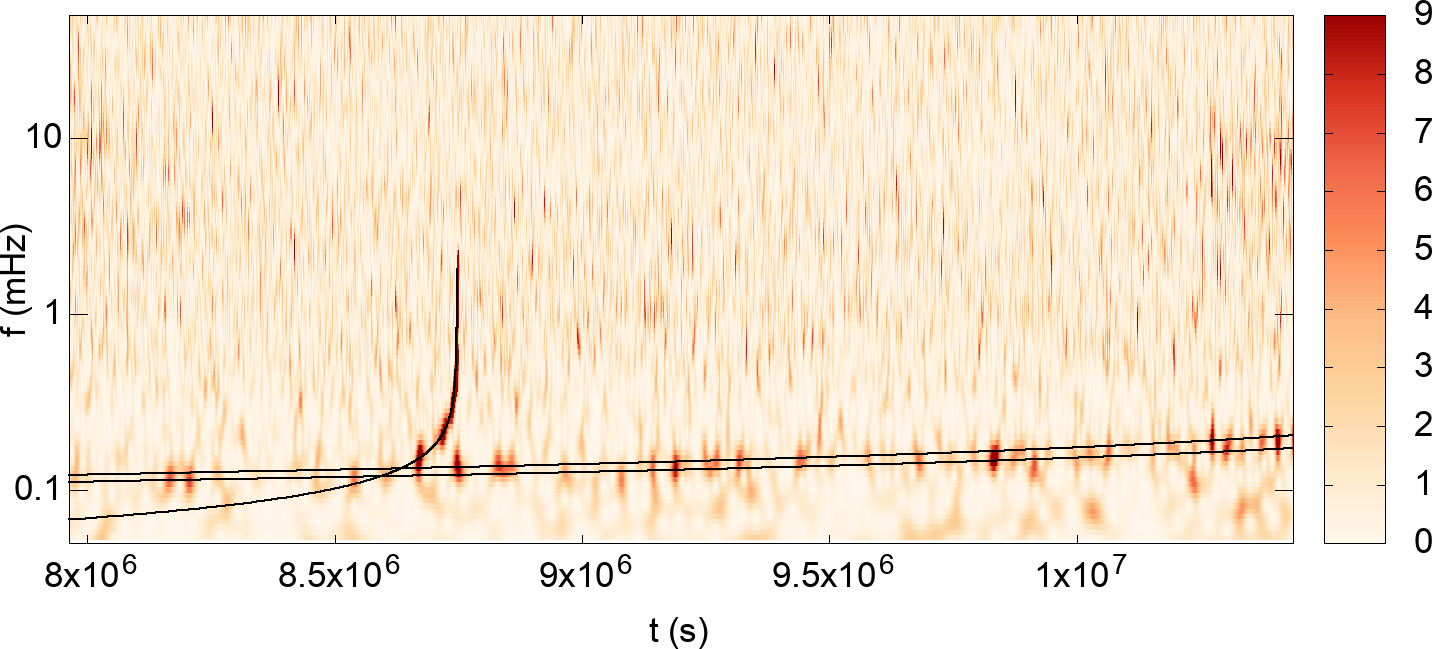} 
\includegraphics[width=0.48\textwidth]{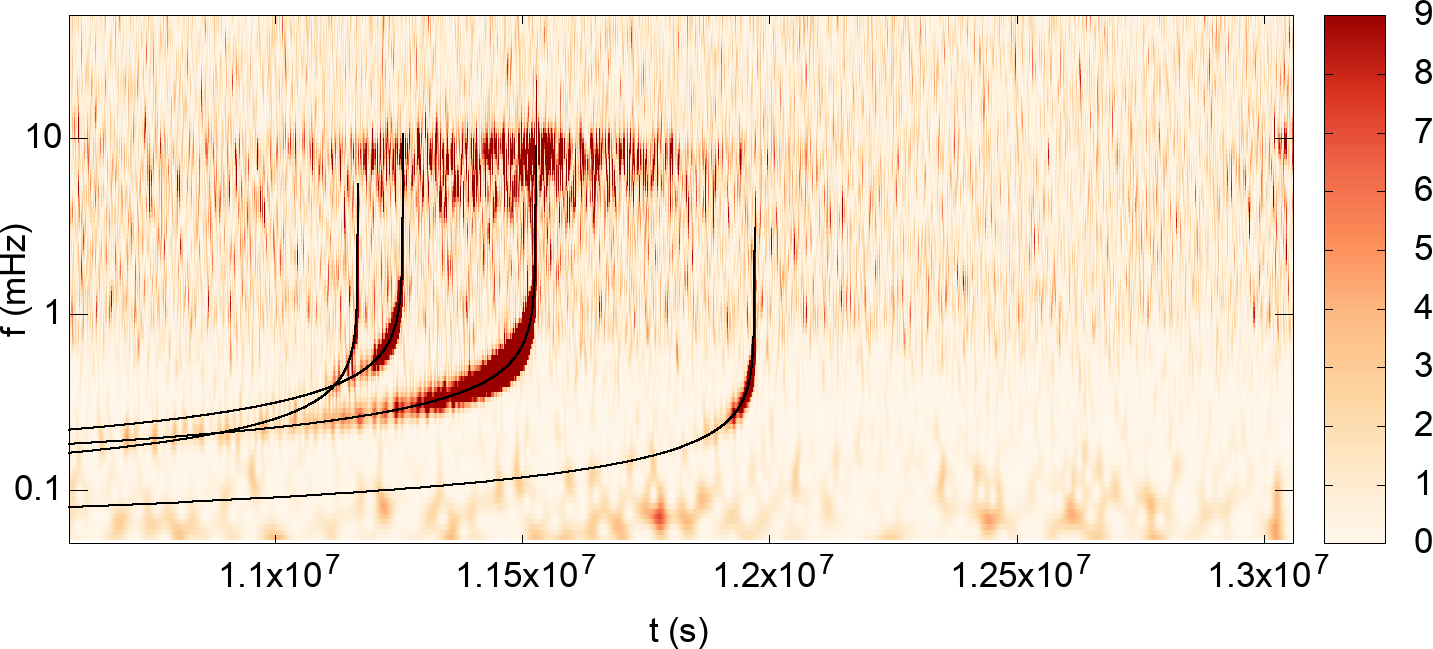} 
\caption{\label{fig:segs68} Spectrograms of the whitened TDI A-channel data from month 4 (upper panel) and month 5 (lower panel) of the LDC {\em Sangria} training data. The black lines indicate the time-frequency tracks of signals found by the search in each segment. Note that two of the sources that merge in month 5 were also picked up prior to merger in month 4.}
\end{figure}

Figure~\ref{fig:segs68} shows time-frequency spectrograms of the whitened TDI A-channel data for months 4 and 5 of the LDC {\em Sangria} training data. Merger signals from four massive black holes are clearly visible in month 5. The merger signal from one binary is visible in month 4, along with fainter tracks from the inspiral phase of two systems that go on to merge in month 5. The search picked up all the mergers, and in addition, picked up the inspiral signal from two of the four systems that went on to merge in month 5. The signal from the other two systems that merge in month 5 were not picked up due to a combination of their higher mass (less time in band) and lower amplitude.

\begin{figure}[htp]
\includegraphics[width=0.48\textwidth]{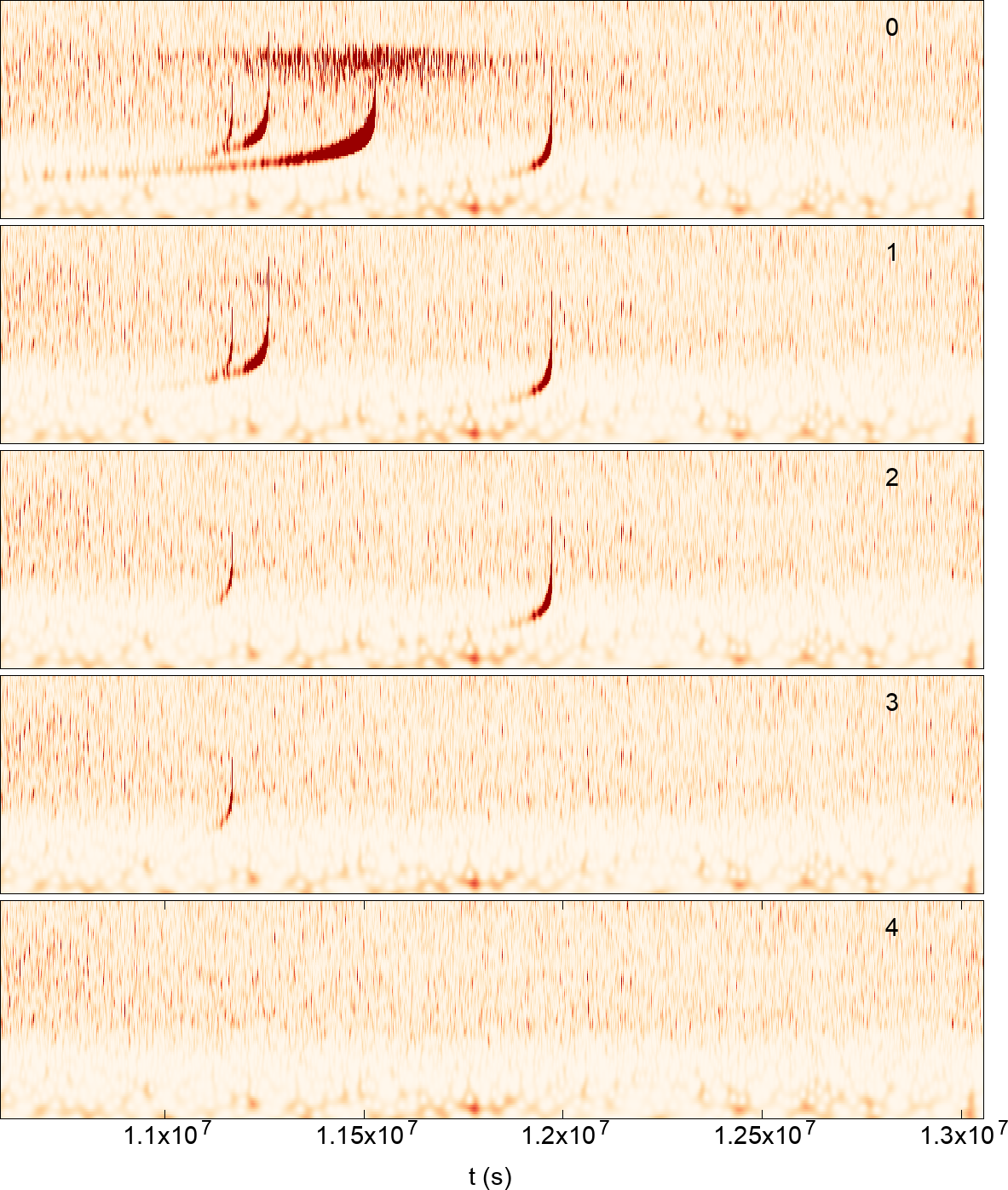} 
\caption{\label{fig:detection} Spectrograms of the whitened TDI A-channel data from month 5 of the LDC {\em Sangria} training data showing the sequential detection and removal of the binary black hole signals.}
\end{figure}

Figure~\ref{fig:detection} illustrates the sequential detection and removal of the binary black hole signals from month 5 of the LDC {\em Sangria} training data set. A sequential solution, rather than a simultaneous solution, is justified here since the overlap, or match, between any pair of signals is tiny. Evaluating the standard expression for the match between signals $h_i,h_j$, $M_{ij} = (h_i|h_j)/\sqrt{(h_i|h_i) (h_j|h_j)}$, where $(a|b)$ is the standard noise weighted inner product of signals $a,b$, we find that the largest match between sources that merge in month 5 occurs for sources 4 and 5, and equals $M_{45} = 8.6 \times 10^{-5}$. One can also consider the phase maximized match, the maximum of which occurs between sources three and four and is equal to
$M^{\rm max}_{34} = 4.9 \times 10^{-3}$. These small match values support using a sequential solution for massive black hole signals, at least for the purposes of low latency analysis.

More generally, the match between two chirping signals can be approximated using the stationary phase approximation, evaluated around the time/frequency where the time frequency tracks $t(f)$ cross:
\begin{equation}\label{match}
(h_i | h_j) \simeq \frac{ 4  \sqrt{2 \pi}  A_i(f_*) A_j(f_*)}{S(f_*)} \cos(\Psi_i(f_*)-\Psi_j(f_*)) \sigma_{f_*} \, ,
\end{equation}
where the critical point $f_*$ is given by the condition $t_i(f_*) = t_j(f_*)$, where $t(f) = \Psi'/(2\pi)$, and $\sigma_{f_*}$ defines the frequency band over which the signals overlap:
\begin{equation}
 \sigma_{f_*}  = \vert \Psi_i^{''}(f_*) - \Psi_j^{''}(f_*) \vert^{-1/2} \, .
\end{equation}
Here $\Psi(f) = \Phi(f) + \phi_p(f)$ is the total phase and $A(f) =  A_p(f) {\cal A}(f)$ is the total amplitude.
During the inspiral phase, and to leading post-Newtonian order, we have
\begin{equation}
 \sigma^2_{f_*}  = \frac{48}{5} \pi^{5/3} f_*^{11/3} \frac{({\cal M}_i {\cal M}_j)^{5/3}}{\vert {\cal M}_j^{5/3}-{\cal M}_i^{5/3} \vert}\, ,
\end{equation}
where ${\cal M} = (m_1 m_2)^{3/5}/(m_1+m_2)^{1/5}$ is the chirp mass. Only sources with very similar chirp masses and merger times will have significant overlaps.

\begin{figure}[htp]
\includegraphics[width=0.48\textwidth]{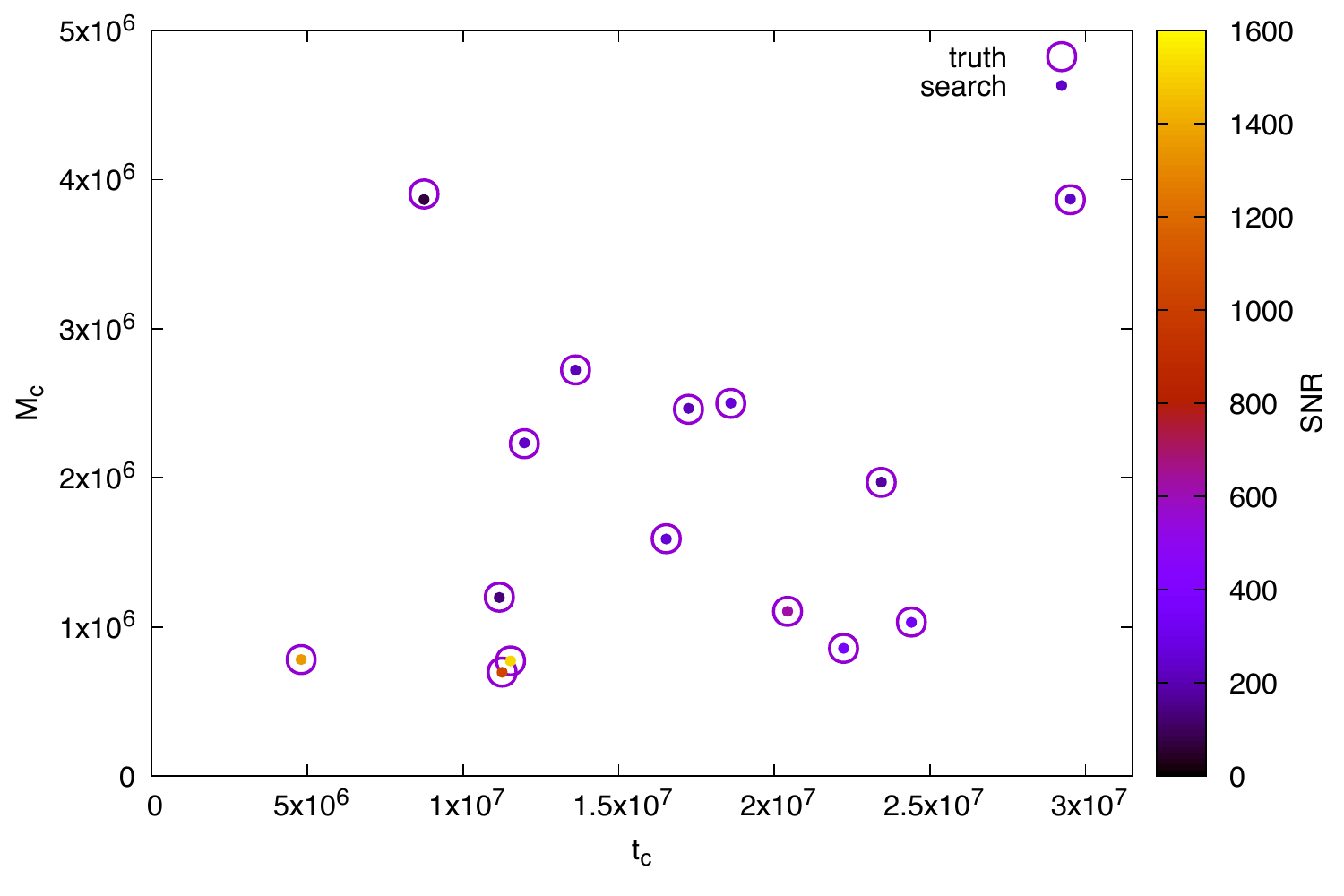} 
\caption{\label{fig:sangria} The chirp masses and merger times for all the sources found by the search of the {\em Sangria} training data. The circles denote the true values, while the colored dots denote the recovered values. The dots are color coded by the SNR. All the simulated sources were accurately recovered.}
\end{figure}

Figure~\ref{fig:sangria} summarizes the output of the initial search of the {\em Sangria} training data set. Many of the sources were found in several of the overlapping one-month data segments. Unique solutions were identified by eliminating solutions with very similar chirp masses and merger times and keeping the solution with the largest signal-to-noise ratio (SNR). The search in each data segment was repeated until the SNR of the candidate signal dropped below a pre-set threshold, in this case ${\rm SNR}_* = 12$. Later, when deciding which candidate solutions to keep, a mass dependent threshold was used: ${\rm SNR}_*(M) = 12 + {\rm ln}(M/ 10^5 M_\odot)$, where $M=m_1+m_2$ is the total mass. The functional form of the threshold was motivated by looking at the SNR distribution of the solutions found when all the true signals had been removed from a segment.

\begin{figure}[H]
\includegraphics[width=0.48\textwidth]{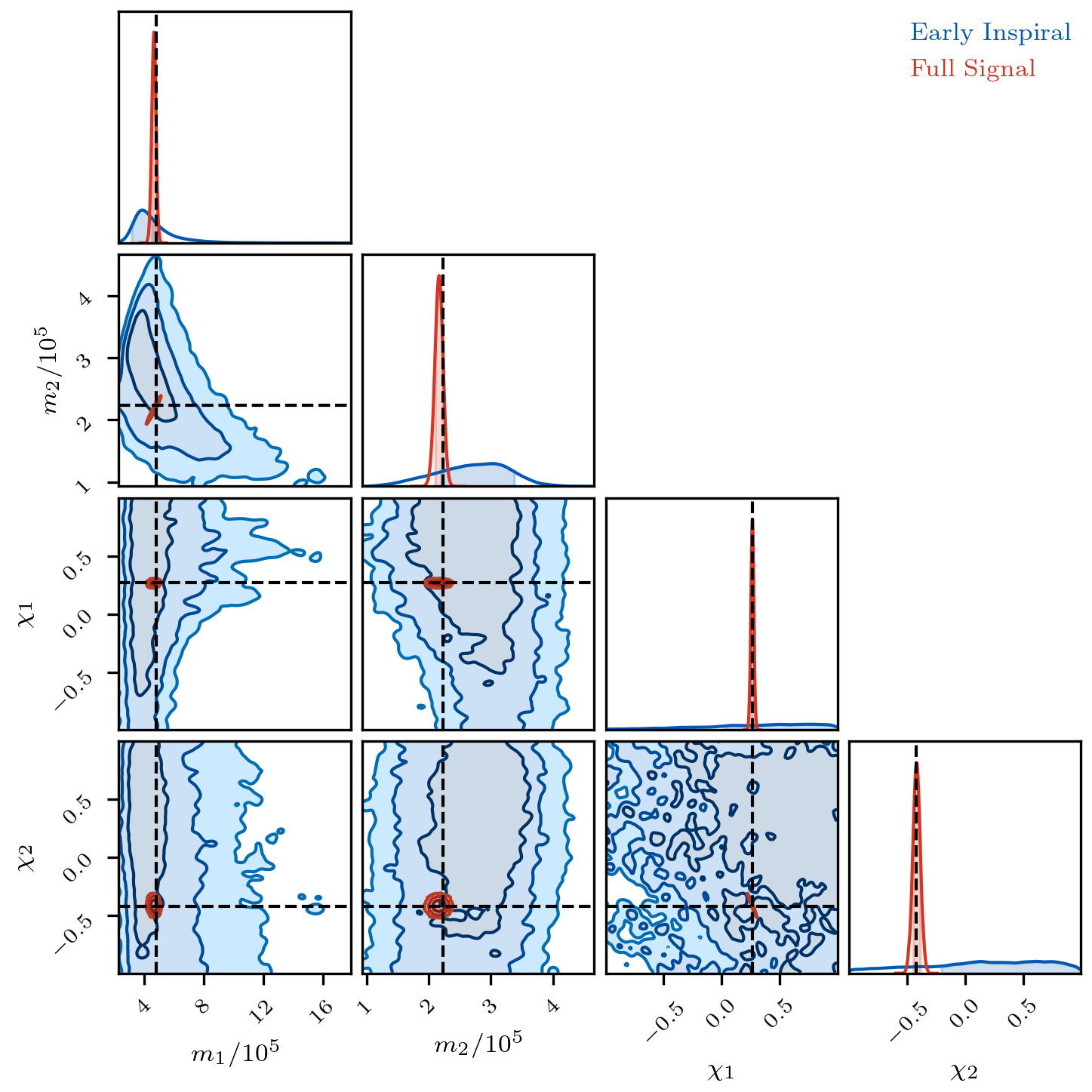} 
\includegraphics[width=0.48\textwidth]{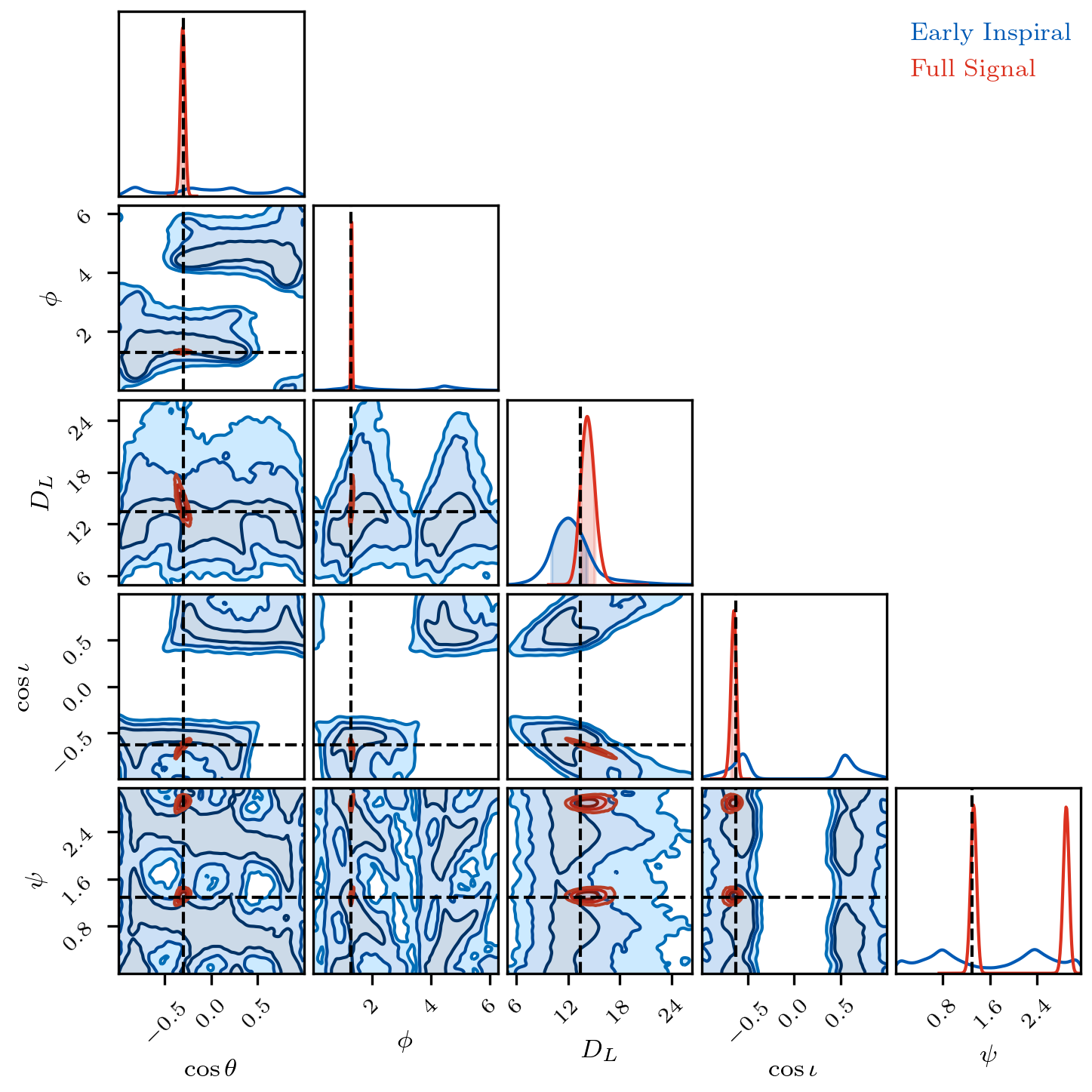} 
\caption{\label{fig:source5} Posterior distributions for source 5 using the full one year data set (in red) and just the month 4 data (in blue) where the inspiral signal was first picked up by the search. The dashed lines indicate the true values of the parameters used to simulated the data. Here the masses are in the source frame, with most of the uncertainty coming from the distance estimate, which is then mapped to a redshift estimate that is used to convert between detector fame and source frame masses.}
\end{figure}

The final stage of the analysis uses a PTMCMC algorithm to map out the posterior distributions for all the source parameters. In this stage of the analysis no maximization is applied to the likelihood. To provide results in low latency a heterodyned likelihood is used~\cite{Cornish:2010kf,Cornish:2021lje}, which for the sources found in the {\em Sangria} data set is typically of order ten thousand times faster than the direct evaluation of the likelihood. The final stage of the analysis is done source-by-source, rather than simultaneously as demanded by the {\em Global Fit}, however all the competing black hole signals are first removed using the best-fit solution from the search, which stops the analysis from wandering off to fit a louder signal.

Figure~\ref{fig:source5} shows marginalized posterior distributions for the parameters of Source 5 using both the full one-year data set, and just the data from month 4 where this sources was first detected. The parameter uncertainties shrink dramatically when the full data set is used. One of the features of the heterodyned likelihood is that its computational cost is independent of the observation time, so the analysis using 12 months of data is just as fast as for 1 month on data. In this case, both took a few hours on a five-year-old 2.9 Ghz Quad-Core MacBook Pro. Note that the parameter uncertainties, especially the sky location and distance, are misleadingly large due to the simplicity of the waveform model being used. More realistic analyses including the effects of additional waveform harmonics and spin precession will significantly improve the parameter resolution~\cite{Lang:2006bsg,Lang:2011je,PhysRevD.76.104016,Porter:2008kn,Marsat:2020rtl}.

\begin{figure}[htp]
\includegraphics[width=0.48\textwidth]{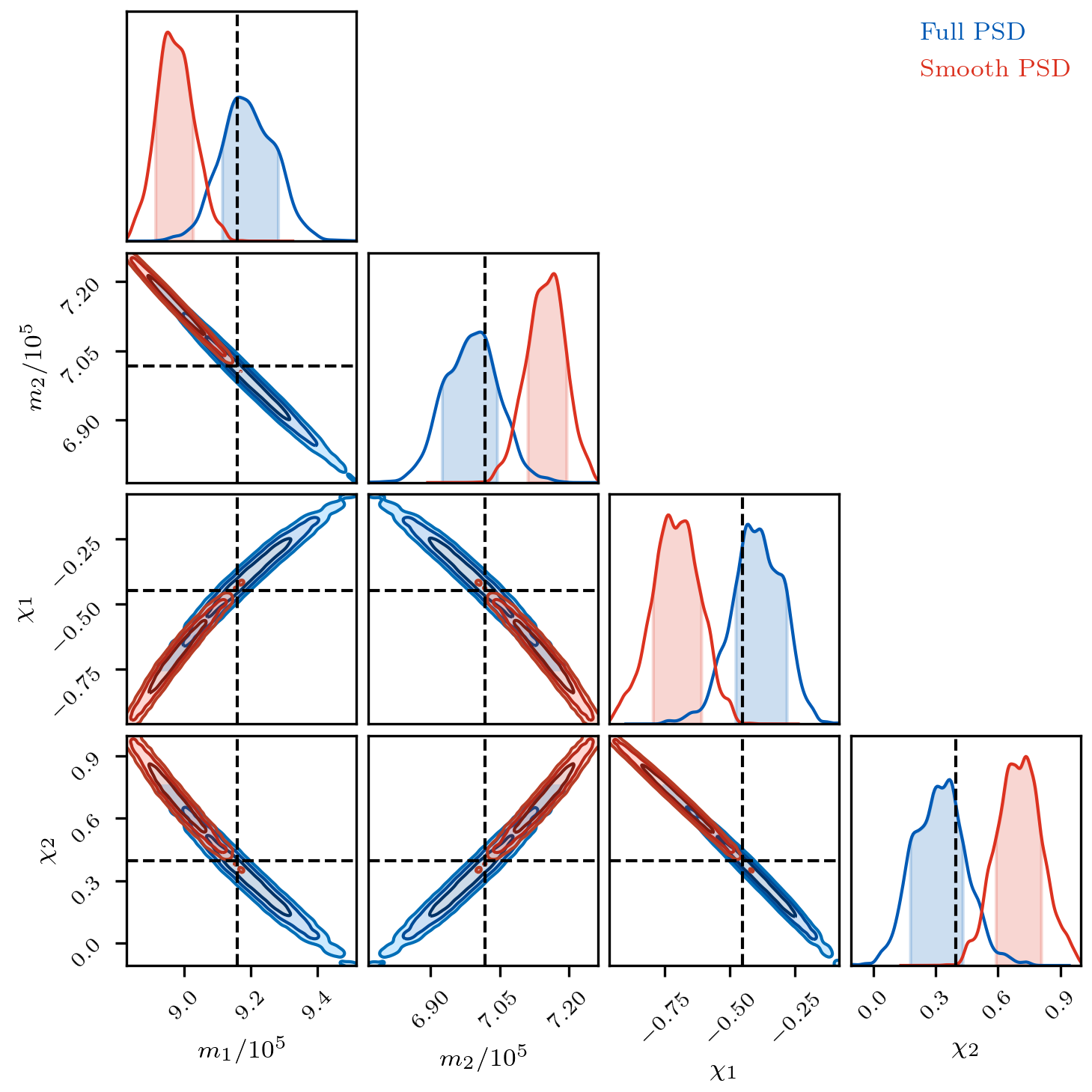} 
\caption{\label{fig:source4} Posterior distributions for the detector frame masses and dimensionless spins for source 4 using the full PSD model (in blue) and the smooth component of the PSD model (in red). The dashed lines indicate the true parameter values used to generate the data. The full PSD model whitens away the loud galactic binaries and returns an unbiased estimate for the masses and spins. The smooth component of the PSD allows the loud galactic binaries to overlap with the black hole signal and cause significant biases in the parameter estimates.}
\end{figure}

While the small overlaps between individual massive black hole signals allow us to get away with a sequential search and source-by-source parameter estimation, the overlaps between the massive black hole signals and the galactic binaries do need to be taken into account. While the overlap between a black hole binary and any one galactic binary is small, there are millions of galactic binaries, and the overlaps combine to have a significant impact. For the numerous quiet galactic binaries, the central limit theorem comes into play and the overlap behaves like Gaussian noise, but for the more sparse population of loud galactic binaries the interaction is non-Gaussian, and can lead to significant biases. In the low latency analysis described here this potential bias is removed by whitening out the signals from the loud binaries. In the {\em Global Fit} the bias is removed by simultaneously analyzing the black hole and galactic binary signals. Figure~\ref{fig:source4} compares the parameter recovery for source 4 using the full PSD estimate, which whitens out the loud galactic binaries, and the smooth PSD estimate, which does not. When the loud galactic binaries are not whitened out the estimates for the component masses and spins are significantly biased.

\section{Summary and Next Steps}

By combining spectral estimation techniques adapted from the analysis of ground based interferometers with a multi-stage stochastic search algorithm, it is possible to perform a robust low latency analysis of simulated LISA data containing multiple overlapping signals. The method is able to provide pre-merger notifications for signals that accumulate sufficient signal-to-noise during the inspiral stage. The search is able to quickly lock onto signals using analytic maximization over extrinsic parameters. The parameter estimates are then refined using a heterodyning technique that dramatically speeds up the calculation of the full likelihood.

The analysis described here was applied to highly simplified simulated data. In reality there will be many additional complications that need to be dealt with, including gaps in the data, non-Gaussian noise transients, non-stationary noise, and more complicated and lower signal-to-noise signals with multiple harmonics, orbital eccentricity and mis-aligned spins. Each of these complications can be dealt with, and will be addressed through future rounds of the LISA Data Challenges. Some ideas about how the complications can be addressed are outlined below.

Systems with unequal masses and/or orbital eccentricity will have multiple harmonics. Since the harmonics are, to a good approximation, orthogonal, they can, for the most part, be dealt with by repeating what was done here for the dominant $(\ell=2,m=2)$ harmonic. For example, each harmonic can be treated separately in the heterodyning procedure. The maximization during the first stage of the search will introduce a separate amplitude and phase for each harmonic in each detector, while the time maximization will be done jointly. The F-statistic likelihood used in the second stage of the search is readily generalize to cover multiple harmonics. Indeed, the F-statistic can be further generalized to cover spin-precessing systems with time evolving inclination and polarization angles~\cite{Fairhurst_2020}. All of this comes at greater computational cost. For example, the F-statistic for precessing systems introduces $4(2\ell +1)$ filters at multipole order $\ell$.

The detection of low signal-to-noise signals, including from systems that are yet to merge, would benefit from an improved maximization scheme during the first stage of the search. In the current implementation a single, roughly month long, segment of data is analyzed, using a constant polarization phase and amplitude. As seen in Figure~\ref{fig:pol}, it is not always a good approximation to treat these quantities as constant for such long durations. Moreover, for low SNR systems it might be necessary to integrate for several months in order to accumulate sufficient SNR to make a detection. One way to overcome both of these problems is to apply the maximization to multiple shorter data segments (say one week in length). But rather than allowing the amplitude and phase to be maximized independently in each segment, which would significantly increase the likelihood even absent a signal, a joint maximization can be applied that enforces a smooth evolution of the polarization amplitude and phase in each TDI channel. The same approach can also be used to detect other low SNR, long lived signals, such as those from stellar original black hole binaries and EMRIs.

The LISA data will suffer from planned and unplanned data gaps~\cite{Baghi:2019eqo} that will impact detection and parameter estimation of massive black hole binaries~\cite{Dey:2021dem}. At a practical level, data gaps complicate noise spectral estimation due to spectral leakage in the Fourier domain~\cite{Baghi:2019eqo} and Discrete Wavelet domain~\cite{Cornish:2020odn}. Various gap-filling techniques have been proposed to mitigate these issues~\cite{Baghi:2019eqo,blelly2021sparse}. Sparse, constrained in-painting in the wavelet domain~\cite{SHEN201626} is a promising technique that can be incorporated in the wavelet de-noising stage of the low latency analysis described here.

The LISA data will likely suffer from non-Gaussian noise transients (glitches), and longer term non-stationarity due to effects such as thermal variations and fluctuations in the solar wind. In the unlikely event that the instrument noise is stationary, the effective noise will be non-stationary due to annual variations in the response to un-resolved galactic binaries, which are modulated by the sweep of the LISA antenna pattern~\cite{Edlund_2005,Adams:2013qma}. In the full {\em Global Fit}, noise transients can be modeled using a variant of the {\tt BayesWave} algorithm~\cite{Cornish:2014kda,Cornish:2020dwh} from ground based gravitational wave astronomy, suitably adapted for space based interferometry~\cite{Robson:2018jly}. Joint Bayesian analysis of binary black hole signals and instrument glitches has been demonstrated for LIGO-Virgo analyses~\cite{Chatziioannou:2021ezd}, but the computational cost may be too high for low latency applications. A LISA specific variant of the glitch-robust {\tt QuickCBC}~\cite{Cornish:2021wxy} algorithm is more appropriate for low latency detection and parameter estimation. For longer term non-stationarity, such as from unresolved galactic binaries, wavelet based methods~\cite{Cornish:2020odn} can be used to infer the dynamic noise spectrum, $S(f,t)$. The heterodyned likelihood naturally incorporates dynamic spectra~\cite{Cornish:2021lje}.

The prototype pipeline described here, enhanced with the developments outlined above, will be ready to analyze real data when they becomes available next decade.

\section*{Acknowledgments}
The author thanks Tyson Littenberg and Stanislav Babak for many informative discussions. This work was supported by NASA LISA foundation Science Grant 80NSSC19K0320, and used data obtained from the LISA Data Challenge {\tt https://lisa-ldc.lal.in2p3.fr}.

\bibliography{refs}

\end{document}